\documentclass[preprint, prb, reprint, twocolumn]{revtex4-2}

\usepackage{
    graphicx,
    orcidlink,
    upgreek
}

\graphicspath{{./images}}

\def\isAmerican{1}
\usepackage{isolects}
\usepackage{variables}
\usepackage{abbreviations}

\newcommand{\NV}{\ensuremath{\mathrm{NV}}}

\newcommand{\nFour}{{\ensuremath{^{14}\text{N}}}}
\newcommand{\nFive}{{\ensuremath{^{15}\text{N}}}}

\begin{document}
\title{Exact Breit-Rabi formulae for the nitrogen-vacancy cent\RE\ in diamond}
\author{Alex~Tritt~\orcidlink{0000-0002-7981-1396}}
\email{alex.tritt@monash.edu}
\author{Lincoln~D.~Turner~\orcidlink{0000-0003-0551-5583}}
\author{Michael~S.~J.~Barson~\orcidlink{0000-0003-0247-5619}}
\affiliation{
School of Physics and Astronomy, Monash University, Victoria 3800, Australia}
\date{\today}

\begin{abstract}
The hyperfine structure that the nitrogen-vacancy (\NV) cent\RE\ in diamond
has with its nitrogen nucleus is of immense importance for quantum sensing and
quantum computing.
Current methods of calculating the \NV\ hyperfine energy eigenvalues and
eigenstates with respect to an axial magnetic field
-- \textit{i.e.}, calculating the \NV\ Breit-Rabi formula --
resort to numerical methods and analytic approximations.
Here we provide an exact, closed-form expression to the \NV\ hyperfine energy
eigenvalues and eigenstates for both the nitrogen-14 and -15 cases.
\end{abstract}

\maketitle

\section{Introduction}

The Breit-Rabi formula is the famous analytic solution of the time-independent
Schr\"odinger equation ($\ham\ket{\psi}=\lambda\ket{\psi}$) for the hyperfine ground state of an alkali atom under
an applied magnetic field~\cite{breit_1931-12-01_physRev-38-2082}.
Importantly, expressions for the energy eigenvalues are exact and valid; both
for low field (anomalous Zeeman effect) and high field (Paschen-Back effect)
regimes, as well as anywhere in
between~\cite[\Section~18.1]{corney_2006_isbn-0-19-921145-0}.
The total number of hyperfine Zeeman levels of a ground-state alkali atom is
$2\times(2\,I + 1)$, with $I$ the nuclear spin quantum number.
While this evaluates to as few as 4 for hydrogen-1,
the second fewest number of levels for an alkali is 6 for
lithium-6 and it grows to as many as 16 for
c\AEE sium-133~\cite{tilley_2002-02-21_nuclearPhysicsA-708-3, khazov_2011-03-29_nuclearDataSheets-112-855}.
Concerningly, Hamiltonians of systems with more than four levels are not
guaranteed to have closed-form expressions for their eigenvalues at all;
at first glance, the task of Breit and Rabi may well have been impossible.
However, in the specific case of ground-state alkalis, the single valence
electron in the s-orbital contributes just two states to the complete atomic
system.
Consequently, expressions for energy eigenvalues of the hyperfine Zeeman system
contain just squares and square roots, making them hyperbolic in magnetic
field~\footnote{
Breit and Rabi's original letter~\cite{breit_1931-12-01_physRev-38-2082} did
not show their working, and subsequent textbooks usually similarly mention the
result without proof.
However, it is left as a guided ``exercise for the reader'' as
\Problem~18.4 of \citet{corney_2006_isbn-0-19-921145-0}.
}.

Now, ninety-five years after Breit and Rabi's letter, researchers of the
nitrogen-vacancy (\NV) col\OU r cent\RE\ in diamond
-- and its applications in quantum technology -- are interested in the form of
its hyperfine Zeeman structure.
The \NV\ is a spin-one defect with a zero-field splitting (ZFS), which means
that it has level crossings and level anticrossings (LACs) at non-zero magnetic
fields (approximately $100~\mathrm{mT}$ and $50~\mathrm{mT}$ respectively for
the electronic ground and excited triplet states).
Probing the hyperfine Zeeman structure of the \NV\ at the LACs has been
fruitful in measuring coupling parameters of the \NV, \textit{viz.}\ the
excited-state hyperfine tensor~\cite{poggiali_2017-05-17_physRevB-95-195308}
and ground state pseudo-Stark tensor~\cite{michl_2019-07-26_nano-lett-19-4904}.
It is expected that similar procedures could also be used to measure the final
two of the six \NV\ stress/strain susceptibility parameters that are yet to be
experimentally quantified~\cite{udvarhelyi_2018-08-02_physRevB-98-075201}.
At the LACs, it is also possible to perform a variety of quantum measurements,
such as creating a $\Lambda-$level structure for
Raman-heterodyne~\cite{holliday_1990-09_opticsLett-15-983, manson_1992_jLumin-53-49},
tuning the spin-resonance frequency close to other systems to
hyperpolar\IS e nearby nuclear spins~\cite{fischer_bulk_2013, jacques_dynamic_2009},
and perform nuclear magnetic resonance (NMR) \cite{jameswood2016-T1NMR} or electron spin resonance (EPR) \cite{armstrongNVNVElectron2010}.
The hyperfine Zeeman level structure of the \NV\ is also important at low
magnetic field.
Knowledge of this structure could be valuable when designing \NV-based quantum
technology for low-field applications, such as zero and ultra-low field (ZULF)
NMR and magnetometry~\cite{omar_2026-03-16_communChem-9-123}.

Unfortunately, the original Breit-Rabi formula
-- valid only for electron-spin-half systems -- cannot be used to determine the
hyperfine Zeeman levels of the electron-spin-one \NV\ cent\RE.
Instead, researchers have used approximate analytic
expressions~\cite{doherty_2012-05-03_physRevB-85-205203, auzinsh_2019-08-14_physRevB-100-075204}
and numerical calculations~\cite{broadway_2016-12-02_physRevApplied-6-064001}
to plot its level structure.
In this manuscript, we show that this is unnecessary;
using reasoning similar to that used by Breit and Rabi, we derive analytic
expressions for the hyperfine Zeeman energy eigenvalues and eigenstates for
the \NV\ cent\RE\ in diamond.

\section{The Zeeman effect on fine structure}
Before discussing the hyperfine Zeeman levels of the \NV, it is worth noting
that an exact, closed-form expression for the energy eigenvalues for the fine
structure of the \NV\ under Zeeman splitting with magnetic fields in arbitrary
directions has only recently been
found~\cite{loenard_2026-01-20_physRevReas-8-013054}.
Before this development, the community has relied on perturbative
solutions~\cite[\Section~IV]{doherty_2012-05-03_physRevB-85-205203}.
In this section, we will add to the work of
\citet{loenard_2026-01-20_physRevReas-8-013054} by finding closed-form
expressions for the energy eigenstates (in terms of the eigenstates of
$\spin_z$, the electronic spin projection operator), as well as the energy
eigenvalues.

\begin{figure}[t]
\includegraphics{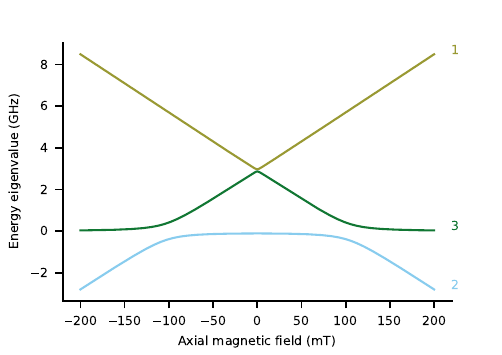}
    \caption{\label{fig:fine-vals}
    Derived energy eigenvalues of the \NV\ electron spin under a magnetic field
    with swept axial component $\field_z$ and a fixed transverse component of
    $\field_x=20\,\mathrm{mT}$.
    Curves are labe\LL ed as per \Equation~\eqref{eq:fine-vals}.
    }
\end{figure}

The fine-structure spin Hamiltonian of the \NV\ under an arbitrary magnetic
field takes the form
\begin{align}
\ham_\mathrm{fine} \,=\,& \zfs\,\spin_z^2 + \fieldV\cdot\gyroM\cdot\spinV,\nonumber\\
    =\,& \zfs\,\spin_z^2 + \gyroP\,\field_x\,\spin_x + \gyroP\,\field_y\,\spin_y + \gyro\,\field_z\,\spin_z,\label{eq:fine-ham}
\end{align}
where $\spinV=(\spin_x,\spin_y,\spin_z)$ is the spin-one vector operator,
$\gyroM = \mathrm{diag}(\gyroP,\gyroP,\gyro)$ is the cylindrical and
almost-isotropic gyromagnetic tensor, $\fieldV=(\field_x,\field_y,\field_z)$ is
the magnetic field, and $\zfs$ is the zero-field splitting (ZFS).
For notational-simplicity, we have set $\hbar=1$.

Before converting \Equation~\eqref{eq:fine-ham} into matrix form, we introduce
the shorthand, $\paulit_{x,y,z}$ for the matrices,
\begin{align}
    \paulit_x \,=&\, \frac1{\sqrt{2}}\begin{pmatrix}0&1&0\\1&0&1\\0&1&0\end{pmatrix}, &
    \paulit_y \,=&\, \frac1{\sqrt{2}}\begin{pmatrix}0&-i&0\\i&0&-i\\0&i&0\end{pmatrix} \nonumber\\
    \paulit_z \,=&\, \begin{pmatrix}1&0&0\\0&0&0\\0&0&-1\end{pmatrix},
\end{align}
and, for later use, $\pauli_{x,y,z}$ for the matrices,
\begin{align}
    \pauli_x \,=&\, \frac12\begin{pmatrix}0&1\\1&0\end{pmatrix}, &
    \pauli_y \,=&\, \frac12\begin{pmatrix}0&-i\\i&0\end{pmatrix}, &
    \pauli_z \,=&\, \frac12\begin{pmatrix}1&0\\0&-1\end{pmatrix}.
\end{align}
Although we use these ``spin matrices'' to literally represent electronic spin
operators $\spin_{x, y, z}$ right now, we refrain from using the symbols
\textit{e.g.} $\mathbf{S}_{x,y,z}$ so that we can reuse them for nuclear spin 
operators $\spinI_{x, y, z}$ and for more-abstract Hamiltonians later in the
manuscript.

In matrix form, \Equation~\eqref{eq:fine-ham} is,
\begin{align}
    \hamM_\mathrm{fine} =&\, \zfs\,\paulit_z^2 + \gyroP\,\field_x\,\paulit_x + \gyroP\,\field_y\,\paulit_y + \gyro\,\field_z\,\paulit_z,\nonumber\\
    =& \begin{pmatrix}
	\zfs + \gyro\,\field_z & \gyroP\frac{\field_x - i\,\field_y}{\sqrt{2}} & 0\\
	\gyroP\frac{\field_x + i\,\field_y}{\sqrt{2}} & 0 & \gyroP\frac{\field_x - i\,\field_y}{\sqrt{2}}\\
	0 & \gyroP\frac{\field_x + i\,\field_y}{\sqrt{2}} & \zfs - \gyro\,\field_z
    \end{pmatrix}.\label{eq:fine-m}
\end{align}
At this point, it is worth asking whether a modern computer algebra system --
like \textsc{Mathematica}~\cite{wolfram_2026_mathematica-15} or
\textsc{SymPy}~\cite{meurer_2017-01-02_peerJCS-3-e103} --
is able find analytic expressions for the eigenvalues of
\Equation~\eqref{eq:fine-m} directly.
We found that they could not.
However, we found both to be helpful in our derivations in the contexts of
finding partial solutions to problems, and checking answers.

We can modify \Equation~\eqref{eq:fine-m} to make its energy eigenvalues
simpler to find.
First, without loss of generality, we can rotate around the $z$ direction into
a coordinate system where the transverse component of the magnetic field is
pointing along the $x$ direction.
This will make matrix real, with real eigenvectors.
Once we have found the eigenvectors of the rotated matrix, we can rotate the
eigenvectors back into the original coordinate system, which will result in the
eigenvectors of \Equation~\eqref{eq:fine-m}.
The rotated matrix is
\begin{align}
    \rotation\,\hamM_\mathrm{fine}\,\rotation^\dagger =& \begin{pmatrix}
    \zfs + \gyro\,\field_z & \gyroP\frac{\field_\perp}{\sqrt{2}} & 0\\
    \gyroP\frac{\field_\perp}{\sqrt{2}} & 0 & \gyroP\frac{\field_\perp}{\sqrt{2}}\\
    0 & \gyroP\frac{\field_\perp}{\sqrt{2}} & \zfs - \gyro\,\field_z
    \end{pmatrix},\label{eq:fine-rot}
\end{align}
with $\field_\perp=\sqrt{\field_x^2 + \field_y^2}$, and the rotation is defined
by $\rotation=\mathrm{diag}(\field_x + i\,\field_y, 0, \field_x - i\,\field_y)/\field_\perp$.

It is instructive to now write \Equation~\eqref{eq:fine-rot} in a
dimensionless form.
We do this by expressing each of its terms in units of $\zfs$; equivalently, by
dividing the entire matrix by $\zfs$.
Resulting in
\begin{align}
\hamM_\mathrm{t3} =&\, \paulit_z^2 + \hor\,\paulit_x + \ver\,\paulit_z\nonumber \\
    =& \begin{pmatrix}
    1 + \ver & \frac\hor{\sqrt{2}} & 0 \\
    \frac\hor{\sqrt{2}} & 0 & \frac\hor{\sqrt{2}} \\
    0 & \frac\hor{\sqrt{2}} & 1 - \ver
    \end{pmatrix},\label{eq:ham-q}
\end{align}
where $\hor = \gyroP\,\field_\perp/\zfs$, $\ver = \gyro\,\field_z/\zfs$.
The eigenvalues of this matrix must have a closed-form solution:
eigenvalues of a matrix are roots of its characteristic polynomial.
Eigenvalues of an $N\times N$ matrix are the roots of its $N$th-order
characteristic polynomial,
$\character_\hamM(\polyvar) = \det\left[\polyvar\,\identity_N - \hamM\right]$,
where $\det[\cdot]$ is the matrix determinant and $\identity_N$ is the
$N\times N$ identity matrix.
For the spin-one system described by $\hamM_\mathrm{t3}$,
$\character_{\hamM_\mathrm{t3}}(\polyvar)$ will be cubic, and so can be solved
using the known expressions of roots for generic cubic equations.

Explicitly, the matrix of \Equation~\eqref{eq:ham-q} has a characteristic
equation of,
\begin{align}
\character_\hamM(\polyvar) =&\, \polyvar^3 - 2\,\polyvar^2 + (-\hor^2 - \ver^2 + 1)\,\polyvar + 2.
\end{align}
If we substitute in $\polyvar = \depvar + 2/3$, then we get a
\emph{depressed cubic}:
a cubic of the form~\cite[\Equation~(1)]{zucker_2008_mathGaz-92-542-264},
\begin{align}
\character'_\hamM(\depvar) =&\, \depvar^3 + 3\,\depp\,\depvar + 2\,\depq,\label{eq:dep}
\end{align}
where, in our case,
\begin{subequations}
\begin{align}
\depp =&\, -1/9 -\hor^2/3 - \ver^2/3,\\
\depq =&\, 1/27 + \hor^2/6 - \ver^2/3.
\end{align}
\end{subequations}
Given that we know that its roots should be real, the roots of
\Equation~\eqref{eq:dep} have a known \emph{trigonometric}
expression~\cite[\Equation~(3)]{zucker_2008_mathGaz-92-542-264},
\begin{align}
\depvar_\rootIndex =&\, 2\,\sqrt{-\depp}\,\cos\left(\frac{\ang + 2\uppi\,\rootIndex}3\right),\quad\text{where}\nonumber\\
    \ang =&\, \arccos\left[\frac{-\depq}{\left(-\depp\right)^{3/2}}\right],
\end{align}
and $j$ is an integer, here we use $j=-1,0,1$.
Substituting in the values for our case yields
\begin{align}
\depvar_\rootIndex =&\, \frac23\,\sqrt{1 + 3\,\hor^2 + 3\,\ver^2}\,\cos\left(\frac{\ang + 2\uppi\,\rootIndex}3\right),\quad\text{and}\nonumber\\
\polyvar_\rootIndex =&\, \frac23\left[1 + \sqrt{1 + 3\,\hor^2 + 3\,\ver^2}\,\cos\left(\frac{\ang + 2\uppi\,\rootIndex}3\right)\right],\quad\text{where}\nonumber\\
    \ang =&\, \arccos\left[-\frac{1  + 9\,\hor^2/2 - 9\,\ver^2}{\left(1 + 3\,\hor^2 + 3\,\ver^2\right)^{3/2}}\right].\label{eq:cubic-sln}
\end{align}

\begin{figure}[t]
\includegraphics{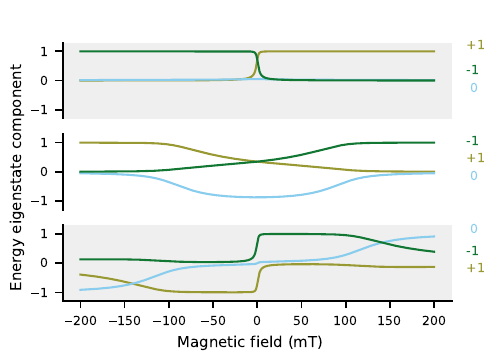}
    \caption{\label{fig:fine-vects}
    Derived coefficients of energy eigenstates of the \NV\ electron spin under
    a magnetic field with swept axial component $\field_z$ and a fixed
    transverse component of $\field_x=20\,\mathrm{mT}$.
    Subplots are states 1 through 3 from top to bottom, as per
    \Equation~\eqref{eq:fine-vects}.
    Coefficients are labe\LL ed by electron magnetic number $\magnetic_S$.
    }
\end{figure}

\textsc{Mathematica} was not able to derive \Equation~\eqref{eq:cubic-sln}.
However, it was able to produce an implicit expression for the eigenvectors of
\Equation~\eqref{eq:ham-q} in terms of the eigenvalues $\polyvar_\rootIndex$.
Explicitly, the unnormal\IS ed eigenstates are,
\begin{align}
    \wavefunctionV_\rootIndex =& \begin{pmatrix}
    2\,\polyvar_\rootIndex\left[\polyvar_\rootIndex + \ver -1\right] - \hor^2\\
    \sqrt{2}\,\hor\left[\polyvar_\rootIndex + \ver - 1\right]\\
    \hor^2
    \end{pmatrix}\label{eq:states-implicit},
\end{align}
for each of the $\rootIndex=-1,0,1$.
Using our expression for the $\polyvar_\rootIndex$ in
\Equation~\eqref{eq:cubic-sln}, the $\wavefunctionV_\rootIndex$ in
\Equation~\eqref{eq:states-implicit} evaluate (ignoring normal\IS ation) to
\begin{align}
\wavefunctionV_\rootIndex=&\, \
    \eV_+\times\bigg[8\,\left(1 + 3\,\hor^2 + 3\,\ver^2\right)\cos^2\left(\frac{\ang + 2\uppi\,\rootIndex}3\right)\nonumber\\
    &\quad+4\left(1 + 3\,\ver\right)\,\sqrt{1 + 3\,\hor^2 + 3\,\ver^2}\,\cos\left(\frac{\ang + 2\uppi\,\rootIndex}3\right)\nonumber\\
    &\quad+12\,\ver-9\,\hor^2 - 4\bigg]\nonumber\\
    &+\eV_0\times\sqrt{2}\,\hor\,\bigg[3\,\ver - 1\nonumber\\
    &\quad+2\,\sqrt{1 + 3\,\hor^2 + 3\,\ver^2}\,\cos\left(\frac{\ang + 2\uppi\,\rootIndex}3\right)\bigg],\nonumber\\
    &+\eV_-\times 9\,\hor^2,\label{eq:states-explicit}
\end{align}
with $\eV_+=(1, 0, 0)$, $\eV_0=(0, 1, 0)$, and $\eV_-=(0, 0, 1)$.
To confirm that these expressions are correct, we multiplied each of the
$\wavefunctionV_\rootIndex$ by $(\polyvar_\rootIndex\,\identity_3 - \hamM)$,
and obtained the zero vector.
To evaluate this algebraically, one has to use the fact that, by construction,
$\cos[(\ang + 2\uppi\,\rootIndex)/3]$ satisfies
\begin{align}
    &3\,\cos\left(\frac{\ang + 2\uppi\,\rootIndex}3\right) - 4\,\cos^3\left(\frac{\ang + 2\uppi\,\rootIndex}3\right)\nonumber\\
    &= \frac{1 + 9\,\hor^2/2 - 9\,\ver^2}{\left(1 + 3\,\hor^2 + 3\,\ver^2\right)^{3/2}}.
\end{align}

Thus, we have derived trigonometric expressions for the fine structure of the
\NV\ under arbitrary magnetic fields;
not only for the eigenvalues as in \citet{loenard_2026-01-20_physRevReas-8-013054}
[\Equation~\eqref{eq:cubic-sln}], but also for eigenstates
[\Equation~\eqref{eq:states-explicit}].
We expand these expressions in \Appendix~\ref{appx-fine}.
We plot the form of the energy eigenvalues in \Figure~\ref{fig:fine-vals}, and the
coefficients of the eigenstates in \Figure~\ref{fig:fine-vects}.
Although eigenvalues look hyperbolic with $\field_z$ close to the LACs, they
are clearly not on a large scale.

\section{A Breit-Rabi formula for nitrogen-15}

\begin{figure}[t]
\includegraphics{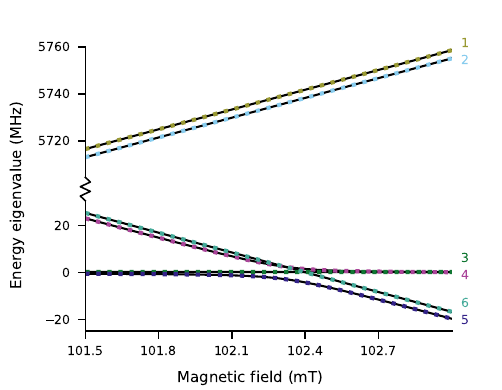}
    \caption{\label{fig:vals-15}
    Comparison between our derived analytic eigenvalues (col\OU red dots) and
    point wise numerical evaluations (black lines) for the \nFive\  cent\RE\
    at the GSLAC.
    Numbers on the right refer to the state labels of
    \Equation~\eqref{eq:15N-sln}.
    Derived analytic curves and numerical evaluations were confirmed to
    be equal to \textsc{float64} precision~\cite{ieeeStd_2019-07-22_ieee-754}.
}
\end{figure}

Nitrogen has two stable isotopes, nitrogen-14 and -15.
Nitrogen-14 is the most abundant, and has a spin-one nucleus, whereas \nFive\
has a spin-half nucleus.
Here the nuclear ($S=1/2$) and electron ($S=1$) spins play opposite roles to the
original Breit-Rabi formula with a single outer electron spin ($S=1/2$) and
larger nuclear spin
(\textit{e.g.}, rubidium-85 and -87 with $I=5/2$ and $I=3/2$ respectively).
Given that there is still a $1/2$ spin in the problem, we can expect the
\nFive\ case to have similar hyperbolic expressions for eigenvalues to that of
the alkalis.
Hence, we derive the Breit-Rabi formula for the \nFive\ vacancy cent\RE\
first.
Further, we use this simpler case to introduce some of the reasoning used in
the more complicated -- yet more common -- \nFour\  case.

The ground state of the \NV\ cent\RE\ with a \nFive\  nucleus, a magnetic
field $\field$ in the $z$ direction \emph{only}, and no strain or electric
fields, has a Hamiltonian of
\begin{align}
\ham_{\nFive} =&\, \zfs\,\spin_z^2
+ \ax\,\spin_x\,\spinI_x + \ax\,\spin_y\,\spinI_y + \az\,\spin_z\,\spinI_z \nonumber\\
    &+ \field\times(\gyro\,\spin_z - \gyroI\,\spinI_z),\label{eq:15N-abstract}
\end{align}
where $\spin_{x,y,z}$ and $\spinI_{x,y,z}$ are the spin-one electron and
spin-half nuclear spin operators, $A_{\parallel,\perp}$ are longitudinal and
transverse hyperfine couplings, and $\gyro$ and $\gyroI$ are the electron and
nuclear gyromagnetic ratios.
We can calculate the matrix form of \Equation~\eqref{eq:15N-abstract} as,
\begin{align}
\hamM_\nFive =&\, \zfs\,\paulit_z^2\otimes\identity_2
+ \ax\,\paulit_x\otimes\pauli_x + \ax\,\paulit_y\otimes\pauli_y \nonumber\\
& + \az\,\paulit_x\otimes\pauli_x
+ \field\times(\gyro\,\paulit_z\otimes\identity_2 - \gyroI\,\identity_3\otimes\pauli_z),\label{eq:15N}
\end{align}
where $\otimes$ is the Kronecker
product~\cite[\Equation~(9.4)]{bengtsson_2017_isbn-978-1-107-02625-4}.

Explicitly writing out the matrix of \Equation~\eqref{eq:15N}, yields,
\begin{widetext}
\begin{align}
    \hamM_\nFive =& \begin{pmatrix}
	\zfs + \frac\az2 + (\gyro - \frac\gyroI2)\,\field & 0 & 0 & 0 & 0 & 0 \\
	0 & \zfs - \frac\az2 + (\gyro + \frac\gyroI2)\,\field & \frac\ax{\sqrt{2}} & 0 & 0 & 0 \\
	0 & \frac\ax{\sqrt{2}} & -\frac{\gyroI\,\field}2 & 0 & 0 & 0 \\
	0 & 0 & 0 & \frac{\gyroI\,\field}2 & \frac\ax{\sqrt{2}} & 0 \\
	0 & 0 & 0 & \frac\ax{\sqrt{2}} & \zfs - \frac\az2 - (\gyro + \frac\gyroI2)\,\field & 0 \\
	0 & 0 & 0 & 0 & 0 & \zfs + \frac\az2 - (\gyro - \frac\gyroI2)\,\field
\end{pmatrix}.
\end{align}
\end{widetext}

\begin{figure}[t]
\includegraphics{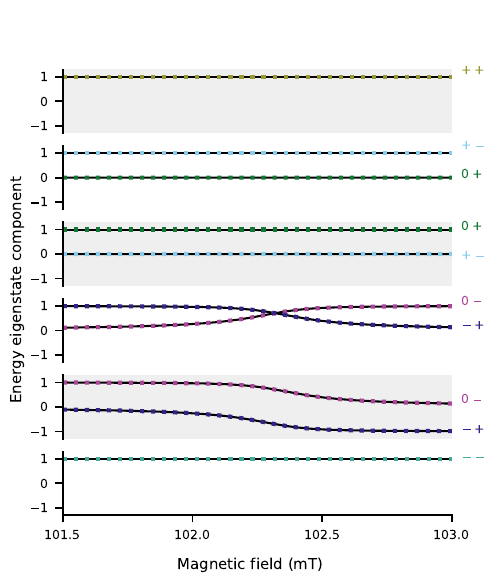}
    \caption{\label{fig:vects-15}
    Comparison between our derived analytic eigenvectors of
    \Equation~\eqref{eq:15N-vec} (col\OU red dots) and point wise numerical
    evaluations (black lines) for the \nFive\  cent\RE\ at the GSLAC.
    Each sub-figure corresponds to a different eigenstate of
    \Equation~\eqref{eq:15N-vec}, in ascending order of $\ket{1}$ at the top
    and $\ket{6}$ on the bottom.
    Each curve corresponds to the coefficient of the
    $\ket{\magnetic_S, \magnetic_I}$ labe\LL ed on the right,
    with only non-zero coefficients are shown.
    Derived analytic coefficients and numerical evaluations were confirmed to
    be equal to \textsc{float64} precision~\cite{ieeeStd_2019-07-22_ieee-754}.
}
\end{figure}

This is obviously a block diagonal matrix.
In fact, each diagonal block has a special property:
that they act on states with the same total magnetic quantum number,
$\magnetic_F = \magnetic_S + \magnetic_I$.
That is to say, a state acted-upon by only one of the blocks will also be an
eigenstate of the hyperfine spin operator $\spinF_z = \spin_z + \spinI_z$,
which has the diagonal matrix representation
$\spinFM_z=\paulit_z\otimes\identity_2 + \identity_3\otimes\pauli_z = \mathrm{diag}(+3/2, +1/2, +1/2, -1/2, -1/2, -3/2)$.
The number $\magnetic_F$ is the eigenvalue of this operator for the state.
This block diagonal form is implied by \Equation~\eqref{eq:15N}, because none
of the terms change $\magnetic_F$:
the only coupling terms (proportional to $x$ or $y$ spin operators) transform
$\magnetic_S$ into $\magnetic_I$, and vice-versa.
Note that this is \emph{not} the case if we introduce electric field and strain
terms into the Hamiltonian.
Methods from this manuscript may not be able to find explicit forms of the
energy eigenvalues for Hamiltonians with such terms, and finding them may
instead require further perturbative or approximation methods.

We can anal\YS e each of the unique diagonal blocks in the Hamiltonian separately.
These are,
\begin{align}
\hamS_{\pm3/2} =&\, \zfs + \frac\az2 \pm \left(\gyro - \frac\gyroI2\right)\,\field,\\
\hamM_{\pm1/2} =& \begin{pmatrix}
\zfs - \frac\az2 \pm (\gyro + \frac\gyroI2)\,\field & \frac\ax{\sqrt{2}}\\
\frac\ax{\sqrt{2}} & \mp\frac{\gyroI\,\field}2
\end{pmatrix}\nonumber\\
    =& \left[\zfs - \frac\az2 \pm \left(\gyro + \gyroI\right)\,\field\right]\pauli_z \nonumber\\
    & + \sqrt{2}\,\ax\,\pauli_x + \left(\zfs - \frac\az2 \pm \gyro\,\field\right)\identity_2,\label{eq:ham-pm12}
\end{align}
where the subscript on the Hamiltonian indicates the value of $\magnetic_F$ for
the subsystem.
The block $\hamS_{\pm3/2}$ acts on the subsystem of
$\ket{\magnetic_S, \magnetic_I} = \ket{\pm1,\pm1/2}$, and
$\hamM_{\pm1/2}$ acts on the subsystem of
$\left(\ket{\pm1,\mp1/2}, \ket{0,\pm1/2}\right)$.
While explicit general algebraic expressions exist for roots of polynomials of
degree four or less, the symmetry groups of polynomial expressions themselves
mean it is a mathematical impossibility for such a general algebraic expression
to exist for roots of polynomials of degree five and
greater~\footnote{
The study of the symmetry groups of polynomial roots is called Galois
theory~\cite{escofier_2001_isbn-0-387-98765-7}.
The fact that there cannot be an algebraic formula for the roots to a
general polynomial of degree five or greater is the Abel-Ruffini theorem
~\cite[\Section~12.5.4]{escofier_2001_isbn-0-387-98765-7}.
}.
This means that while there may not have been an algebraic expression for the
eigenvalues and eigenstates for a generic eight-level Hamiltonian, by splitting
it up into Hamiltonians of subsystems of at most two levels, we can be sure
that such an algebraic expression exists.

Note that \Equation~\eqref{eq:ham-pm12} has a standard form of,
\begin{align}
\hamM_\mathrm{t2} =&\, \ver\,\pauli_z + \hor\,\pauli_x + \bias\,\identity_2,\label{eq:pauli}
\end{align}
parameter\IS ed by $\ver$, $\hor$ and $\bias$.
This form of Hamiltonian is well known to have eigenvalues
of~\cite[\Equations~(3.56) and (3.69)]{sakurai_2020-09-17_modernQuantumMechanics-isbn-978-1-108-47322-4},
\begin{align}
\polyvar_\pm = \left(\bias \pm \sqrt{\hor^2 + \ver^2}\right)/2,
\end{align}
with (unnormal\IS ed) eigenstates,
\begin{align}
    \wavefunctionV_\pm =& \begin{pmatrix}
	\ver \pm \sqrt{\hor^2 + \ver^2}\\
	+\hor
    \end{pmatrix}.
\end{align}
Hence, the full set of six energy levels are succinctly,
\begin{align}
    \energy =& \begin{cases}
	\zfs + \frac\az2 \pm \left(\gyro - \frac\gyroI2\right)\,\field,& \magnetic_F=\pm3/2\\ \\
\left(\frac\zfs2 - \frac\az4 \pm^1 \frac{\gyro\,\field}2\right)\\
	\pm^2 \sqrt{\frac{\ax^2}2 + \left(\frac\zfs2 - \frac\az4 \pm^1 \frac{\gyro + \gyroI}2\,\field\right)^2}, & \magnetic_F=\pm^1 1/2
\label{eq:lite-15-energies}
    \end{cases}
\end{align}
with eigenstates proportional to
\begin{align}
    &\ket{\wavefunction} = \nonumber\\
    &\begin{cases}
	\ket{\pm1,\pm1/2}, & \magnetic_F = \pm\frac32\\ \\
    \ket{\pm^1 1,\mp^1 1/2}\Bigg( \left[\zfs - \frac{\az}2 \pm^1 \left(\gyro + \gyroI\right)\field\right]\\
    \pm^2 \sqrt{2\,\ax^2 + \left[\zfs - \frac{\az}2 \pm^1 \left(\gyro + \gyroI\right)\field\right]^2}\Bigg)\\
	+ \ket{0,\pm^1 1/2}\,\sqrt{2}\,\ax, & \magnetic_F = \pm^1 \frac12
    \end{cases}\label{eq:lite-15-states}
\end{align}
where $\pm^1$ and $\pm^2$ should be evaluated independently.

We write the states and levels out in full in \Appendix~\ref{appx-15}.
We plot the evaluations of the eigenvalues in \Equation~\eqref{eq:15N-sln} in
\Figure~\ref{fig:vals-15}, and the coefficients of the  eigenstates in
\Equation~\eqref{eq:15N-vec} in \Figure~\ref{fig:vects-15}.

The level anticrossing of the \nFive\ vacancy cent\RE\ is at the magnetic field
$\field_\mathrm{LAC}$ that minim\IS es the radicand (argument) of
the square roots in \Equations~(\ref{eq:15N-sln}b-e).
That is, when,
\begin{align}\label{eq:lac-15}
    \field_\mathrm{LAC} =&\,\frac{\zfs - \az/2}{\gyro + \gyroI}.
\end{align}

When we use the ground-state (gs) world averages for the \nFive V cent\RE\
parameters~\cite{typical-nv}
    $\zfs^\mathrm{gs}=2.8694(5)\,\mathrm{GHz}$,
    $\az^\mathrm{gs}=3.03331(13)\,\mathrm{MHz}$,
    $\gyro^\mathrm{gs}=28.0331(28)\,\mathrm{GHz/T}$, and
    $\gyroI=-4.315255(21)\,\mathrm{MHz/T}$,
\Equation~\eqref{eq:lac-15} predicts a ground state level anticrossing (GSLAC)
at $\field_\mathrm{GSLAC} = 102.318(19)\,\mathrm{mT}$.
When we use the excited-state (es) world averages for the \nFive V cent\RE\
parameters~\cite{typical-nv}
$\zfs^\mathrm{es}=1.42(6)\,\mathrm{GHz}$,
$\az^\mathrm{es}=-61(6)\,\mathrm{MHz}$,
$\gyro^\mathrm{es}=28.25(26)\,\mathrm{GHz/T}$, and
$\gyroI=-4.315255(21)\,\mathrm{MHz/T}$,
\Equation~\eqref{eq:lac-15} predicts an excited state level anticrossing
(ESLAC) at $\field_\mathrm{ESLAC} = 51.3(2.2)\,\mathrm{mT}$.
\break

\section{A Breit-Rabi formula for nitrogen-14}

The \nFour\ vacancy cent\RE\ under an axial magnetic field has a Hamiltonian
of the form,
\begin{align}
\ham_\nFour =&\, \zfs\,\spin_z^2 + \qua\,\spinI_z^2
    + \ax\,\spin_x\,\spinI_x + \ax\,\spin_y\,\spinI_y \nonumber\\
    & + \ax\,\spin_z\,\spinI_z + \field\,(\gyro\,\spin_z - \gyroI\,\spinI_z).\label{eq:14N-abstract}
\end{align}
It differs from the form of the \nFive\ vacancy cent\RE\
\Equation~\eqref{eq:15N-abstract} in that the \nFour\ nucleus has a nuclear
qudrupole moment $\qua$, owing to it being spin-one rather than spin-half.
Note that, though given the same symbols and playing the same roles, the
values for the nuclear and hyperfine parameters $\gyroI$, $\ax$, and $\az$ have
different values for the two isotopes.
The matrix form of the Hamiltonian is
\begin{align}
\hamM_\nFour =&\, \zfs\,\paulit_z^2\otimes\identity_3 + \qua\,\identity_3\otimes\paulit_z^2 \nonumber\\
    &+ \ax\,\paulit_x\otimes\paulit_x + \ax\,\paulit_y\otimes\paulit_y
 + \az\,\paulit_x\otimes\paulit_x\nonumber\\
    &+ \field\,(\gyro\,\paulit_z\otimes\identity_3 - \gyroI\,\identity_3\otimes\paulit_z).\label{eq:14N}
\end{align}

When written explicitly, we get
\begin{widetext}
\begin{align}
    &\hamM_\nFour = \nonumber\\
    &\begin{tiny}\begin{pmatrix}
    \az + \zfs + \qua + \field\left(\gyro - \gyroI\right) & 0 & 0 & 0 & 0 & 0 & 0 & 0 & 0\\
    0 & \zfs + \gyro\,\field & 0 & \ax &  0 & 0 & 0 & 0 & 0\\
    0 & 0 & \zfs + \qua - \az + \field(\gyro + \gyroI) & 0 & \ax & 0 & 0 & 0 & 0\\
    0 & \ax & 0 & \qua - \gyroI\,\field & 0 & 0 & 0 & 0 & 0 \\
    0 & 0 & \ax & 0 & 0 & 0 & \ax & 0 & 0\\
    0 & 0 & 0 & 0 & 0 & \qua + \gyroI\,\field & 0 & \ax & 0 \\
    0 & 0 & 0 & 0 & \ax & 0 & \zfs + \qua - \az - \field(\gyro + \gyroI) & 0 & 0\\
    0 & 0 & 0 & 0 &  0 & \ax & 0 & \zfs - \gyro\,\field & 0\\
    0 & 0 & 0 & 0 & 0 & 0 & 0 & 0 & \az + \zfs + \qua - \field\left(\gyro - \gyroI\right)
\end{pmatrix}\end{tiny}.
\end{align}
Unlike the \nFive\  matrix, it is not immediately in block diagonal form.
However, it can be transformed into one simply by reordering the entries with
a particular permutation matrix $\reorder$.
The reordered block diagonal matrix evaluates to,
\begin{align}
&\reorder\,\hamM_\nFour\,\reorder = \nonumber\\
&\begin{tiny}\begin{pmatrix}
\az + \zfs + \qua + \field\left(\gyro - \gyroI\right) & 0 & 0 & 0 & 0 & 0 & 0 & 0 & 0\\
0 & \zfs + \gyro\,\field & \ax & 0 & 0 & 0 & 0 & 0 & 0\\
0 & \ax & \qua - \gyroI\,\field & 0 & 0 & 0 & 0 & 0 & 0\\
0 & 0 & 0 & \zfs + \qua - \az + \field(\gyro + \gyroI) & \ax & 0 & 0 & 0 & 0 \\
0 & 0 & 0 & \ax & 0 & \ax & 0 & 0 & 0\\
0 & 0 & 0 & 0 & \ax & \zfs + \qua - \az - \field(\gyro + \gyroI) & 0 & 0 & 0\\
0 & 0 & 0 & 0 & 0 & 0 & \qua + \gyroI\,\field & \ax & 0\\
0 & 0 & 0 & 0 & 0 & 0 & \ax & \zfs - \gyro\,\field & 0\\
0 & 0 & 0 & 0 & 0 & 0 & 0 & 0 & \az + \zfs + \qua - \field\left(\gyro - \gyroI\right)
\end{pmatrix}\end{tiny},\label{eq:14N-matr}
\end{align}
\end{widetext}
where,
\begin{align}
    \reorder =& \begin{pmatrix}
1 & 0 & 0 & 0 & 0 & 0 & 0 & 0 & 0 \\
0 & 1 & 0 & 0 & 0 & 0 & 0 & 0 & 0 \\
0 & 0 & 0 & 1 & 0 & 0 & 0 & 0 & 0 \\
0 & 0 & 1 & 0 & 0 & 0 & 0 & 0 & 0 \\
0 & 0 & 0 & 0 & 1 & 0 & 0 & 0 & 0 \\
0 & 0 & 0 & 0 & 0 & 0 & 1 & 0 & 0 \\
0 & 0 & 0 & 0 & 0 & 1 & 0 & 0 & 0 \\
0 & 0 & 0 & 0 & 0 & 0 & 0 & 1 & 0 \\
0 & 0 & 0 & 0 & 0 & 0 & 0 & 0 & 1
\end{pmatrix}.
\end{align}
Similarly to the \nFive\  case, these blocks act on states with the same
hyperfine magnetic number $\magnetic_F$.
Explicitly, $\reorder\,\spinFM_z\,\reorder = \reorder(\paulit_z\otimes\identity_3 + \identity_3\otimes\paulit_z)\reorder = \mathrm{diag}(+2, +1, +1, 0, 0, 0, -1, -1, -2)$.
Writing each block separately,
\begin{subequations}
\begin{align}
\hamS_{\pm2} =&\, \az + \zfs + \qua \pm \field\left(\gyro - \gyroI\right),\label{eq:ham-2}\\
\hamM_{\pm1} =&\, \begin{pmatrix}
\zfs \pm \gyro\,\field & \ax \\
\ax & \qua \mp \gyroI\,\field
\end{pmatrix}, \nonumber \\
=& \left[\zfs - \qua \pm \field\left(\gyro + \gyroI\right)\right]\,\pauli_z\nonumber\\
&+2\,\ax\,\pauli_x + \left[\zfs + \qua \pm \field\left(\gyro - \gyroI\right)\right]\,\identity_2,\label{eq:ham-1}\\
    \hamM_0 =& \begin{tiny}\begin{pmatrix}
\zfs + \qua - \az + \field(\gyro + \gyroI) & \ax & 0\\
\ax & 0 & \ax\\
0 & \ax & \zfs + \qua - \az - \field(\gyro + \gyroI)
    \end{pmatrix}\end{tiny},\nonumber\\
=& \left(\zfs + \qua - \az\right)\,\paulit_z^2 + \sqrt{2}\,\ax\,\paulit_x + \field(\gyro + \gyroI)\,\paulit_z,\label{eq:ham-0}
\end{align}
\end{subequations}
where, again, the subscript of the Hamiltonian corresponds to the magnetic
$\magnetic_F$ number of the states it acts on.
The block $\hamS_{\pm2}$ acts on the state
$\ket{\magnetic_S,\magnetic_I} = \ket{\pm1, \pm1}$, the block $\hamM_{\pm1}$ on
state vector $\left(\ket{\pm1,0}, \ket{0, \pm1}\right)$, and the block
$\hamM_0$ on state vector
$\left(\ket{+1, -1}, \ket{0, 0}, \ket{-1, +1}\right)$.

As with the \nFive\ case, while a generic nine-level Hamiltonian might
not have closed-form expressions for its energy eigenvalues, by splitting it
into Hamiltonians of at most three levels, we can be sure that an algebraic
expression does exist.
In fact, we have already found such expressions:
$\hamM_{\pm1}$ have the same form as \Equation~\eqref{eq:pauli},
and $\hamM_0$ has the same form as \Equation~\eqref{eq:ham-q}.
We can therefore use our previous results to write the energy eigenvalues of
the \nFour\ cent\RE.
\begin{align}
\energy =& \begin{cases}
    \az + \zfs + \qua \pm \field\left(\gyro - \gyroI\right), \quad\quad\quad\quad \magnetic_F=\pm2 \\ \\
    \frac{\zfs + \qua}2 \pm^1 \frac{\gyro - \gyroI}2\field\\
    \pm^2{\frac12}\sqrt{4\,\ax^2 + \left[\zfs - \qua \pm^1 \left(\gyro + \gyroI\right)\field\right]^2},\\
    \quad\quad\quad\quad\quad\quad\quad\quad\quad\quad\quad\quad\quad\quad\quad\quad \magnetic_F = \pm^1 1\\ \\
    \frac23\left(\zfs + \qua - \az\right) + \frac23\cos\left(\frac{\ang + 2\uppi\,\rootIndex}3\right)\\
    \times\sqrt{\left(\zfs + \qua - \az\right)^2 + 6\,\ax^2 + 3\,(\gyro + \gyroI)^2\,\field^2},\\
    \quad\quad\quad\quad\quad\quad\quad\quad\quad\quad\quad\quad\quad\quad\quad\quad \magnetic_F = 0
    \end{cases}\\
    \quad\mathrm{with}\nonumber\\
    \ang =& \arccos\Biggl(\left[\az - \zfs - \qua\right]\nonumber\\
    &\times\frac{\left(\zfs + \qua - \az\right)^2 + 9\,\ax^2 - 9\,(\gyro + \gyroI)^2\,\field^2}{\left[\left(\zfs + \qua - \az\right)^2 + 6\,\ax^2 + 3\,(\gyro + \gyroI)^2\,\field^2\right]^{3/2}}\Biggl).
\label{eq:lite-14-energies}
\end{align}

\begin{figure}[t]
\includegraphics{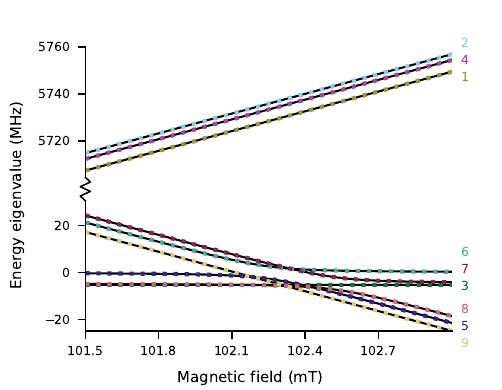}
    \caption{\label{fig:vals-14}
    Comparison between our derived analytic eigenvalues (col\OU red dots) and
    point wise numerical evaluations (black lines) for the \nFour\  cent\RE\
    at the GSLAC.
    Numbers on the right refer to the state labels of
    \Equation~\eqref{eq:14N-sln}.
    Errors (bottom) are the absolute difference of the two evaluations, shown
    on a log scale.
    Derived analytic curves and numerical evaluations were confirmed to
    be equal to \textsc{float64} precision~\cite{ieeeStd_2019-07-22_ieee-754}.
}
\end{figure}

\begin{figure}[t]
\includegraphics{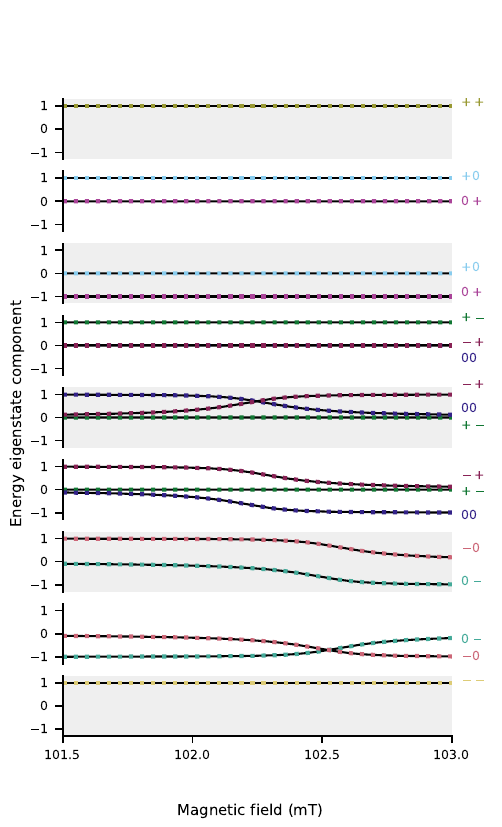}
    \caption{\label{fig:vects-14}
    Comparison between our derived analytic eigenvectors of
    \Equation~\eqref{eq:14N-vec} (col\OU red dots) and point wise numerical
    evaluations (black lines) for the \nFour\  cent\RE\ at the GSLAC.
    Each sub-figure corresponds to a different eigenstate of
    \Equation~\eqref{eq:14N-vec}, in ascending order of $\ket{1}$ at the top
    and $\ket{9}$ on the bottom.
    Each curve corresponds to the coefficient of the
    $\ket{\magnetic_S, \magnetic_I}$ labe\LL ed on the right,
    with only non-zero coefficients are shown.
    Derived analytic coefficients and numerical evaluations were confirmed to
    be equal to \textsc{float64} precision~\cite{ieeeStd_2019-07-22_ieee-754}.
}
\end{figure}

We write the states and levels out in full in \Appendix~\ref{appx-14}.
We plot the evaluations of the eigenvalues in \Equation~\eqref{eq:14N-sln} in
\Figure~\ref{fig:vals-14}, and the eigenstates in \Equation~\eqref{eq:14N-vec}
in \Figure~\ref{fig:vects-14}.
We predict the level anticrossings for this system later in the manuscript.

\section{Quadratic approximations}

\begin{figure}[t]
\includegraphics{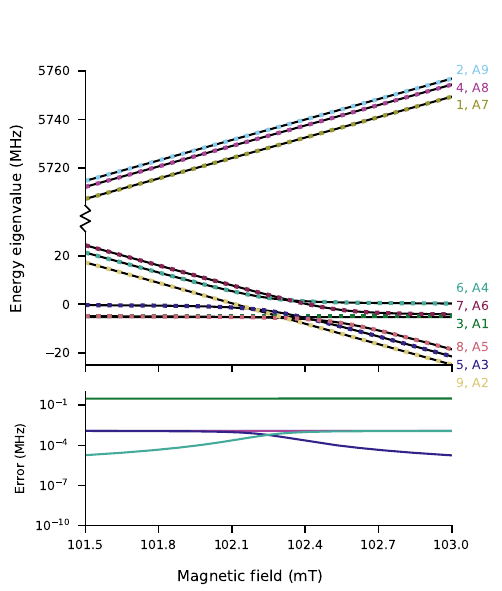}
    \caption{\label{fig:vals-auzinish}
    Comparison between the approximate analytic eigenvalues
    of \citet{auzinsh_2019-08-14_physRevB-100-075204}
    -- which we derive from our exact formula in
    \Equation~\eqref{eq:quadratic-sln} --
    (col\OU red dots) and point wise numerical evaluations (black lines) for
    the \nFour\ cent\RE\ at the GSLAC.
    Numbers on the right refer to the state labels of
    our \Equation~\eqref{eq:14N-sln} (left) and \Equation~(4) of
    \citet{auzinsh_2019-08-14_physRevB-100-075204}.
    Errors (bottom) are the absolute difference of the two evaluations, shown
    on a log scale.
    These errors are lower than $1~\mathrm{MHz}$, indicating agreement between
    expressions, but much higher than numerical precision, showing some
    inaccuracy.
}
\end{figure}

\citet{auzinsh_2019-08-14_physRevB-100-075204} derives approximate hyperbolic
expressions to the \nFour\  vacancy cent\RE, with their \Equation~(4)
being their expression for eigenvalues and \Equation~(5) the same for
eigenstates~\cite{fn-gyro}.
They both mostly agree with the numerical results
(and hence our exact results), at least in the domain of applicability around
the GSLAC.

Here we also show how the quadratic expressions of
\citet{auzinsh_2019-08-14_physRevB-100-075204} can be derived from our exact
expressions.
Our purpose of showing this derivation is to give some intuition of how these
non-standard trigonometric expressions match-up with the understanding the \NV\
community already has with standard quadratic expressions.
We start with the general Hamiltonian, and define
$\depP=-9\,\depp = 1 + 3\,\hor^2 + 3\,\ver^2$ and
$\depQ=27\,\depq=1 + 9/2\,\hor^2 - 9\,\ver^2$ for brevity.
We can expand the general eigenvalue expression for $\rootIndex=\pm1$ to,
\begin{align}
\polyvar_\pm =&\, \frac23\left(1 + \sqrt{\depP}\cos(\ang/3\pm2\uppi/3)\right),\nonumber\\
    =&\,\frac13\left[2 - \sqrt{\depP}\left(\cos(\ang/3)\pm\sqrt3\,\sin(\ang/3)\right)\right].
\end{align}

The level anticrossings occur when $\ver\approx1$.
In this case, $-\depQ/\depP^{3/2}\approx1$, meaning that
$\ang\approx\arccos(-1)=0$ is small.
We can then Taylor-expand for small $\ang$ as,
\begin{align}
\polyvar_\pm =&\, \frac13\left[(2 - \sqrt{\depP}) \mp\sqrt{\frac{\depP}3}\ang + \frac{\sqrt{\depP}}{18}\ang^2 \right] + \order(\ang^3).
\end{align}

We now have to expand $\ang=\arccos(\argu)$ with
$\argu=-\depQ/\depP^{3/2}\approx1$.
When $\argu\approx1$, then the derivative of $\arccos(\argu)$ blows-up to
infinity, and a Taylor expansion cannot be taken.
However, a \emph{Puiseux expansion}
-- a series in terms of $\argu$ to rational powers -- can be used.
Intuitively, the fact that $\cos(\alpha)=1 - \alpha^2/2 + \order(\alpha^4)$ has
a quadratic form for small $\alpha$, means that $\arccos(\argu)$ has the form
of a square root function of $\argu$ (a fractional power) when
$\argu\approx1$.
Specifically, the Puiseux expansion of $\arccos$ at $1$
is~\cite[\Equation~(4.24.2)]{olver_2010_isbn-978-0-521-14063-8},
\begin{align}
\arccos(\argu) =&\, \sqrt{2\,(1 - \argu)}\times\big[1 + \order(1 - \argu)\big].
\end{align}
Using this we have,
\begin{align}
\polyvar_\pm =&\, \frac13\left[2 - \frac89\sqrt{\depP} + \frac{\sqrt{\depP}\argu}9\mp\sqrt{\frac23(1 - \argu)\depP}\right]\nonumber\\
    &+ \order(\hor^3 + \verr^3),\label{eq:before-taylor}
\end{align}
where $\verr$ is the expansion of $\ver$ around 1: $\ver = 1 + \verr$.
We now substitute $\argu=-\depQ/\depP^{3/2}\approx1$,
\begin{align}
\polyvar_\pm =&\, \frac13\left[2 - \frac89\sqrt{\depP} + \frac{\depQ}{9\depP}\mp\sqrt{\frac23\left(\depP + \frac{\depQ}{\sqrt{\depP}}\right)}\right]\nonumber\\
&+ \order(\hor^3 + \verr^3).
\end{align}
We can express $\depP$ and $\depQ$ in terms of $\hor$ and $\verr$ as
\begin{subequations}\label{eq:pq-zeta}
\begin{align}
\depQ =&\, 9/2\,\hor^2 - 9\,\verr^2 - 18\,\verr - 8,\\
\depP =&\, 4 + 3\,\hor^2 + 3\,\verr^2 + 6\,\verr,\nonumber\\
=&\, 4\left(1 + \frac34\,\hor^2 + \frac34\,\verr^2 + \frac32\,\verr \right),
\end{align}
\end{subequations}
meaning that functions of $\depP$ can be expanded as,
\begin{subequations}\label{eq:p-taylor}
\begin{align}
\frac1\depP =&\, \frac14\,\bigg[1 - \left(\frac34\,\hor^2 + \frac34\,\verr^2 + \frac32\,\verr\right) \nonumber\\
    &+ \left(\frac34\,\hor^2 + \frac34\,\verr^2 + \frac32\,\verr\right)^2\bigg] + \order(\hor^3 + \verr^3),\\
\sqrt{\depP} =&\, 2\,\bigg[1 + \frac12\,\left(\frac34\,\hor^2 + \frac34\,\verr^2 + \frac32\,\verr\right) \nonumber\\
    &-\frac18\,\left(\frac34\,\hor^2 + \frac34\,\verr^2 + \frac32\,\verr\right)^2\bigg] + \order(\hor^3 + \verr^3),\\
\frac1{\sqrt{\depP}} =&\, \frac12\,\bigg[1 - \frac12\,\left(\frac34\,\hor^2 + \frac34\,\verr^2 + \frac32\,\verr\right) \nonumber\\
    &+ \frac38\left(\frac34\,\hor^2 + \frac34\,\verr^2 + \frac32\,\verr\right)^2\bigg] + \order(\hor^3 + \verr^3).
\end{align}
\end{subequations}
Substituting \Equations~\eqref{eq:pq-zeta} and \eqref{eq:p-taylor} into
\Equation~\eqref{eq:before-taylor} finally gives,
\begin{align}
    \polyvar_\pm =&\, \frac12\left[-\verr\mp\sqrt{2\,\hor^2 + \verr^2}\right] + \order(\hor^2 + \verr^3)
    .\label{eq:quadratic-sln}
\end{align}
In the \nFour\ cent\RE\ expression (after scaling of the whole $\polyvar_\pm$
expression to dimensional units), we have
$\verr \mapsto [\left(\gyro + \gyroI\right)\,\field - \zfs - \qua + \az]$ and
$\hor \mapsto (\sqrt{2}\,\ax)$.
Substituting these into \Equation~\eqref{eq:quadratic-sln}, we recover
exactly~\cite{fn-gyro} the expression for energies 3 and 4 as defined by
\citet{auzinsh_2019-08-14_physRevB-100-075204} in their \Equations~(4c) and
(4d).
We plot these approximate expressions in
\Figure~\ref{fig:vals-auzinish-corrected}.

\section{Predicted level anticrossings}

\begin{figure}[t]
\includegraphics{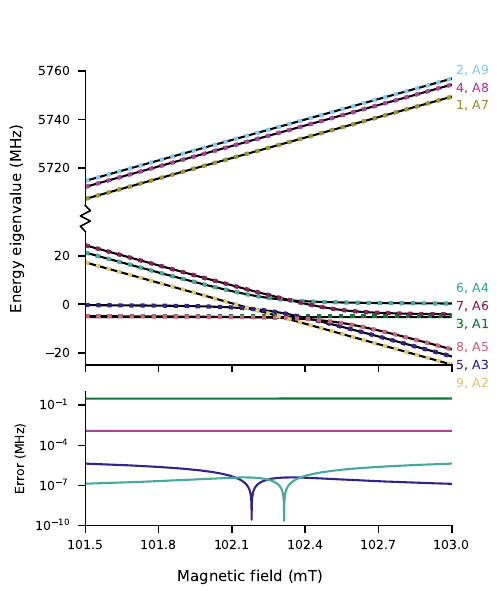}
    \caption{\label{fig:vals-auzinish-corrected}
    Comparison between the approximate analytic eigenvalues
    of \Equation~\eqref{eq:corrected} (col\OU red dots)
    and point wise numerical evaluations (black lines) for the \nFour\ 
    cent\RE\ at the GSLAC.
    Numbers on the right refer to the state labels of
    our \Equation~\eqref{eq:14N-sln} (left) and \Equation~(4) of
    \citet{auzinsh_2019-08-14_physRevB-100-075204}.
    Errors (bottom) are the absolute difference of the two evaluations, shown
    on a log scale.
    The errors for energy levels 5 and 6 are smaller than those in
    \Figure~\ref{fig:vals-auzinish}, indicating that our derived corrections to
    the vertical and horizontal shifts to the level anticrossing are correct.
}
\end{figure}

Note that the \Equation~\eqref{eq:quadratic-sln} is exact to quadratic order in
$\verr$, but only to linear order in $\hor$.
Including the second-order term in $\hor$ gives,
\begin{align}
    \polyvar_\pm =&\, \frac12\left[-\left(\verr + \hor^2/4\right)\mp\sqrt{2\,\hor^2 + \verr^2}\right] + \order(\hor^3 + \verr^3).
\end{align}
Such an $\hor^2$-dependent shift in energy level could be due to a combination
of an incorrect assumption that the level anticrossing is exactly at
$\verr = 0\implies\ver = 1$, or a vertical shift in the energy levels.
We will now show that it is due to both.

The exact magnetic field of the level anticrossing of the three-level subsystem can
be most clearly defined as where a pair of eigenstates have the same magnitude coefficients when expressed 
in the $\ket{m_S,m_I}$ basis.
In other words, it is location where the lines cross in the middle plot of 
\Figure~\ref{fig:vects-14}.                                                     
In this case, from \Equation~\eqref{eq:states-implicit}, we have,
\begin{align}
    0 =&\,\ver - 1 - \hor/\sqrt{2} + \polyvar_+.
\end{align}
Using \Equation~\eqref{eq:before-taylor} we can approximate this as,
\begin{align}
    \sqrt{\frac2{27}(1 - \argu)\depP} =&\, \ver - \frac13 - \frac{\hor}{\sqrt{2}} - \frac89\sqrt{\depP} + \frac{\sqrt{\depP}\argu}9 \nonumber\\
    &+ \order(\hor^3 + \verr^3).
\end{align}
Squaring both sides, using the series expansions of
\Equation~\eqref{eq:p-taylor}, and solving for $\verr$, we find,
\begin{align}
    \verr_\mathrm{LAC} =&\, \frac{\hor^2}4,&
    \ver_\mathrm{LAC} =&\, 1 + \frac{\hor^2}4.\label{eq:lac-0}
\end{align}

This suggests that a better quadratic approximation would be,
\begin{align}
    \polyvar_\pm =&\, \frac12\left[\left(1 + \frac{\hor^2}4\right)- \ver \mp\sqrt{2\,\hor^2 + \left(\ver - \left[1 + \frac{\hor^2}4\right]\right)^2}\right]\nonumber\\
    &- \frac{\hor^2}4 + \order\left[\hor^3 + \left(\ver - \left[1 + \frac{\hor^2}4\right]\right)^3\right].\label{eq:corrected}
\end{align}
We plot this corrected approximation in
\Figure~\ref{fig:vals-auzinish-corrected}.
The errors of the corrected approximation from the numerical evaluation are
indeed smaller that those of the uncorrected approximation, confirming the
merit of the above derivations.

\Equation~\eqref{eq:lac-0}, as well as the quadratic expressions for the
$\magnetic_F=\pm1$ sublevel, imply that the \nFour\  system has level
anticrossings at,
\begin{subequations}
\begin{align}
\field_\mathrm{LAC} =&\, \frac{\zfs - \qua}{\gyro + \gyroI},\quad\mathrm{and}\label{eq:14N-lac-1}\\
    \field_\mathrm{LAC} =&\, \frac{\left(\zfs + \qua - \az\right)^2 + \ax^2/2}{\left(\zfs + \qua - \az\right)\times\left(\gyro + \gyroI\right)}\label{eq:14N-lac-2}.
\end{align}
\end{subequations}
When we use the ground-state (gs) world averages for the \nFour V cent\RE\
parameters~\cite{typical-nv}
$\zfs^\mathrm{gs}=2.8694(5)\,\mathrm{GHz}$,
$\qua=-4.94575(3)\,\mathrm{MHz}$,
$\ax^\mathrm{gs}=-2.6328(12)\,\mathrm{MHz}$,
$\az^\mathrm{gs}=-2.164690(9)\,\mathrm{MHz}$,
$\gyro^\mathrm{gs}=28.0331(28)\,\mathrm{GHz/T}$, and
$\gyroI=3.076272(15)\,\mathrm{MHz/T}$,
\Equation~\eqref{eq:14N-lac-1} predicts a GSLAC
of the $\magnetic_F=\pm1$ levels at
$\field_\mathrm{GSLAC} = 102.522(19)\,\mathrm{mT}$, and
\Equation~\eqref{eq:14N-lac-2} predicts a GSLAC
of the $\magnetic_F=0$ levels at
$\field_\mathrm{GSLAC} = 102.247(21)\,\mathrm{mT}$.
When we use the excited-state (es) world averages for the \nFour V cent\RE\
parameters~\cite{typical-nv}
$\zfs^\mathrm{es}=1.42(6)\,\mathrm{GHz}$,
$\qua=-4.94575(3)\,\mathrm{MHz}$,
$\ax^\mathrm{es}=-23(3)\,\mathrm{MHz}$,
$\az^\mathrm{es}=40\,\mathrm{MHz}$,
$\gyro^\mathrm{es}=28.25(26)\,\mathrm{GHz/T}$, and
$\gyroI=3.076272(15)\,\mathrm{MHz/T}$,
\Equation~\eqref{eq:14N-lac-1} predicts an ESLAC
of the $\magnetic_F=\pm1$ levels at
$\field_\mathrm{ESLAC} = 50.4(2.2)\,\mathrm{mT}$ and
\Equation~\eqref{eq:14N-lac-2} predicts an ESLAC
of the $\magnetic_F=0$ levels at
$\field_\mathrm{ESLAC} = 48.7(2.2)\,\mathrm{mT}$.
In particular, we find that $\magnetic_F=0$ predictions are well within
uncertainties of what would be predicted if we used the original assumption of
$\ver=1$ and ignored the $\ax$ term in \Equation~\eqref{eq:14N-lac-2}.

\section{Conclusion}

We have derived exact Breit-Rabi formulae for the energy eigenvalues and
eigenstates of both the \nFour\  cent\RE\ and \nFive\  cent\RE\ in
diamond.
Numerical experiments and an analytic derivation show that these expressions
agree with previous analytic approximations in their domain of applicability.
These expressions are useful when simple linear expressions breakdown.
In particular at the level-anticrossings near zero magnetic field, applicable to ZULF magnetometry, 
and at the ESLAC and GSLAC where interesting spin-resonance and spectroscopy measurements are performed.

\section*{\ACKNOWLEDGEMENTS}

This work was funded by the National Intelligence and Security Discovery
Research Grant (NISDRG) number NI240100144, provided by the Australian Office
of National Intelligence (ONI).
M.S.J.B.\ is supported by an Australian Research Council (ARC) Early Career
Industry Fellowship (IE24) number IE240100023.
We also thank Leon B.\ Miller, Josh P.\ Duff, Taylor J.\ Christie, and
Kris Helmerson for helpful discussions, with special thanks to JPD and TJC for
proofreading this manuscript.

\clearpage
\begin{appendix}
\begin{widetext}

\section{Explicit expressions for the fine structure of the NV under arbitrary magnetic fields}\label{appx-fine}

For the fine-structure Zeeman Hamiltonian of \Equation~\eqref{eq:fine-ham},
the energy eigenvalues of found in \Equation~\eqref{eq:cubic-sln} expand to

\begin{subequations}\label{eq:fine-vals}
\begin{align}
    \energy_1 =&\, \frac23\left[\zfs + \sqrt{\zfs^2 + 3\,\gyroP^2\,\field_x^2 + 3\,\gyroP^2\,\field_y^2 + 3\,\gyro^2\,\field_z^2}
    \cos\left(\frac{\ang}3\right)\right],\\
    \energy_2 =&\, \frac23\left[\zfs + \sqrt{\zfs^2 + 3\,\gyroP^2\,\field_x^2 + 3\,\gyroP^2\,\field_y^2 + 3\,\gyro^2\,\field_z^2}
    \cos\left(\frac{\ang + 2\uppi}3\right)\right],\\
    \energy_3 =&\, \frac23\left[\zfs + \sqrt{\zfs^2 + 3\,\gyroP^2\,\field_x^2 + 3\,\gyroP^2\,\field_y^2 + 3\,\gyro^2\,\field_z^2}
    \cos\left(\frac{\ang - 2\uppi}3\right)\right],
\end{align}
\end{subequations}

and the unnormal\IS ed eigenstates of \Equation~\eqref{eq:states-explicit}
expand to

\begin{subequations}\label{eq:fine-vects}
\begin{align}
\ket{1} =&\,
    \ket{\magnetic_S=+1}\times\Bigg[8\,\left(\zfs^2 + 3\,\gyroP^2\,\field_x^2 + 3\,\gyroP^2\,\field_y^2 + 3\,\gyro^2\,\field_z^2\right)
    \cos^2\left(\frac{\ang}3\right)\nonumber\\
    &\quad\quad\quad\quad\quad\quad\quad+4\,\left(\zfs + 3\,\gyro\,\field_z\right)\sqrt{\zfs^2 + 3\,\gyroP^2\,\field_x^2 + 3\,\gyroP^2\,\field_y^2 + 3\,\gyro^2\,\field_z^2}\cos\left(\frac{\ang}3\right)\nonumber\\
    &\quad\quad\quad\quad\quad\quad\quad+12\,\zfs\,\gyro\field_z - 9\,\gyroP^2\,\field_x^2 - 9\,\gyroP^2\,\field_y^2 - 4\,\zfs^2
    \Bigg]\left(\field_x + i\,\field_y\right) \nonumber\\
    &+\ket{\magnetic_S=0}\times3\,\sqrt{2}\,\gyroP\left(\field_x^2+\field_y^2\right)\left[3\,\gyro\,\field_z - \zfs
    +2\,\sqrt{\zfs^2 + 3\,\gyroP^2\,\field_x^2 + 3\,\gyroP^2\,\field_y^2 + 3\,\gyro^2\,\field_z^2}\cos\left(\frac{\ang}3\right)\right]\nonumber\\
    &+ \ket{\magnetic_S=-1}\times9\,\gyroP^2\left(\field_x^2 + \field_y^2\right)\left(\field_x - i\,\field_y\right),\\
    \nonumber\\
\ket{2} =&\,
    \ket{\magnetic_S=+1}\times\Bigg[8\,\left(\zfs^2 + 3\,\gyroP^2\,\field_x^2 + 3\,\gyroP^2\,\field_y^2 + 3\,\gyro^2\,\field_z^2\right)
    \cos^2\left(\frac{\ang + 2\uppi\,\rootIndex}3\right)\nonumber\\
    &\quad\quad\quad\quad\quad\quad\quad+4\,\left(\zfs + 3\,\gyro\,\field_z\right)\sqrt{\zfs^2 + 3\,\gyroP^2\,\field_x^2 + 3\,\gyroP^2\,\field_y^2 + 3\,\gyro^2\,\field_z^2}\cos\left(\frac{\ang + 2\uppi\,\rootIndex}3\right)\nonumber\\
    &\quad\quad\quad\quad\quad\quad\quad+12\,\zfs\,\gyro\field_z - 9\,\gyroP^2\,\field_x^2 - 9\,\gyroP^2\,\field_y^2 - 4\,\zfs^2
    \Bigg]\left(\field_x + i\,\field_y\right) \nonumber\\
    &+\ket{\magnetic_S=0}\times3\,\sqrt{2}\,\gyroP\left(\field_x^2+\field_y^2\right)\left[3\,\gyro\,\field_z - \zfs
    +2\,\sqrt{\zfs^2 + 3\,\gyroP^2\,\field_x^2 + 3\,\gyroP^2\,\field_y^2 + 3\,\gyro^2\,\field_z^2}\cos\left(\frac{\ang + 2\uppi\,\rootIndex}3\right)\right]\nonumber\\
    &+ \ket{\magnetic_S=-1}\times9\,\gyroP^2\left(\field_x^2 + \field_y^2\right)\left(\field_x - i\,\field_y\right),\\
    \nonumber\\
\ket{3} =&\,
    \ket{\magnetic_S=+1}\times\Bigg[8\,\left(\zfs^2 + 3\,\gyroP^2\,\field_x^2 + 3\,\gyroP^2\,\field_y^2 + 3\,\gyro^2\,\field_z^2\right)
    \cos^2\left(\frac{\ang - 2\uppi\,\rootIndex}3\right)\nonumber\\
    &\quad\quad\quad\quad\quad\quad\quad+4\,\left(\zfs + 3\,\gyro\,\field_z\right)\sqrt{\zfs^2 + 3\,\gyroP^2\,\field_x^2 + 3\,\gyroP^2\,\field_y^2 + 3\,\gyro^2\,\field_z^2}\cos\left(\frac{\ang - 2\uppi\,\rootIndex}3\right)\nonumber\\
    &\quad\quad\quad\quad\quad\quad\quad+12\,\zfs\,\gyro\field_z - 9\,\gyroP^2\,\field_x^2 - 9\,\gyroP^2\,\field_y^2 - 4\,\zfs^2
    \Bigg]\left(\field_x + i\,\field_y\right) \nonumber\\
    &+\ket{\magnetic_S=0}\times3\,\sqrt{2}\,\gyroP\left(\field_x^2+\field_y^2\right)\left[3\,\gyro\,\field_z - \zfs
    +2\,\sqrt{\zfs^2 + 3\,\gyroP^2\,\field_x^2 + 3\,\gyroP^2\,\field_y^2 + 3\,\gyro^2\,\field_z^2}\cos\left(\frac{\ang - 2\uppi\,\rootIndex}3\right)\right]\nonumber\\
    &+ \ket{\magnetic_S=-1}\times9\,\gyroP^2\left(\field_x^2 + \field_y^2\right)\left(\field_x - i\,\field_y\right),
\end{align}
\end{subequations}
with
\begin{align}
\ang =&\, \arccos\left[-\zfs\,\frac{\zfs^2 + 9\,\gyroP^2\,\field_x^2/2 + 9\,\gyroP^2\,\field_y^2/2 - 9\,\gyro^2\,\field_z^2}
    {\left(\zfs^2 + 3\,\gyroP^2\,\field_x^2 + 3\,\gyroP^2\,\field_y^2 + 3\,\gyro^2\,\field_z^2\right)^{3/2}} \right].
\end{align}
\clearpage

\section{Explicit Breit Rabi formula for \nFive\ cent\RE}\label{appx-15}

For the \nFive-cent\RE\ Zeeman Hamiltonian of \Equation~\eqref{eq:15N-abstract},
the energy eigenvalues of found in \Equation~\eqref{eq:lite-15-energies} expand to

\begin{subequations}\label{eq:15N-sln}
\begin{align}
\energy_1 =&\, \zfs + \az/2 + \left(\gyro - \gyroI/2\right)\field,\\
\energy_2 =&\, \frac12\,\left[\zfs - \az/2 + \gyro\,\field
+ \sqrt{2\,\ax^2 + \left[\zfs - \az/2 + \left(\gyro + \gyroI\right)\,\field\right]^2}\right],\\
\energy_3 =&\, \frac12\,\left[\zfs - \az/2 + \gyro\,\field
- \sqrt{2\,\ax^2 + \left[\zfs - \az/2 + \left(\gyro + \gyroI\right)\,\field\right]^2}\right],\\
\energy_4 =&\, \frac12\,\left[\zfs - \az/2 - \gyro\,\field
+ \sqrt{2\,\ax^2 + \left[\zfs - \az/2 - \left(\gyro + \gyroI\right)\,\field\right]^2}\right],\\
\energy_5 =&\, \frac12\,\left[\zfs - \az/2 - \gyro\,\field
- \sqrt{2\,\ax^2 + \left[\zfs - \az/2 - \left(\gyro + \gyroI\right)\,\field\right]^2}\right],\\
\energy_6 =&\, \zfs + \az/2 - \left(\gyro - \gyroI/2\right)\field.
\end{align}
\end{subequations}

and the unnormal\IS ed eigenstates of \Equation~\eqref{eq:lite-15-states}
expand to

\begin{subequations}\label{eq:15N-vec}
\begin{align}
\ket{1} =&\, \ket{+1,+1/2},\\
\ket{2} =&\, \ket{+1,-1/2}\times\left( \left[\zfs - \frac{\az}2 + \left(\gyro + \gyroI\right)\field\right]
+ \sqrt{2\,\ax^2 + \left[\zfs - \frac{\az}2 + \left(\gyro + \gyroI\right)\field\right]^2}\right)\nonumber\\
    &+ \ket{0,+1/2}\times\sqrt{2}\,\ax,\\
\ket{3} =&\, \ket{+1,-1/2}\times\left( \left[\zfs - \frac{\az}2 + \left(\gyro + \gyroI\right)\field\right]
- \sqrt{2\,\ax^2 + \left[\zfs - \frac{\az}2 + \left(\gyro + \gyroI\right)\field\right]^2}\right)\nonumber\\
&+ \ket{0,+1/2}\times\sqrt{2}\,\ax,\\
\ket{4} =&\, \ket{-1,+1/2}\times\left( \left[\zfs - \frac{\az}2 - \left(\gyro + \gyroI\right)\field\right]
+ \sqrt{2\,\ax^2 + \left[\zfs - \frac{\az}2 - \left(\gyro + \gyroI\right)\field\right]^2}\right)\nonumber\\
    &+ \ket{0,-1/2}\times\sqrt{2}\,\ax,\\
\ket{5} =&\, \ket{-1,+1/2}\times\left( \left[\zfs - \frac{\az}2 - \left(\gyro + \gyroI\right)\field\right]
- \sqrt{2\,\ax^2 + \left[\zfs - \frac{\az}2 - \left(\gyro + \gyroI\right)\field\right]^2}\right)\nonumber\\
    &+ \ket{0,-1/2}\times\sqrt{2}\,\ax,\\
\ket{6} =&\, \ket{-1, -1/2}.
\end{align}
\end{subequations}
\clearpage

\section{Explicit Breit Rabi formula for \nFour\ cent\RE}\label{appx-14}

For the \nFour-cent\RE\ Zeeman Hamiltonian of \Equation~\eqref{eq:14N-abstract},
the energy eigenvalues of found in \Equation~\eqref{eq:lite-14-energies} expand to

\begin{subequations}\label{eq:14N-sln}
\begin{align}
\energy_1 =&\, \az + \zfs + \qua + \left(\gyro - \gyroI\right)\field,\\
\energy_2 =&\, \frac12\left[\zfs + \qua + \left(\gyro - \gyroI\right)\field
+\sqrt{4\,\ax^2 + \left[\zfs - \qua + \left(\gyro + \gyroI\right)\field\right]^2}\right],\\
\energy_3 =&\, \frac12\left[\zfs + \qua + \left(\gyro - \gyroI\right)\field
-\sqrt{4\,\ax^2 + \left[\zfs - \qua + \left(\gyro + \gyroI\right)\field\right]^2}\right],\\
\energy_4=&\, \frac23\left[\zfs + \qua - \az + \cos\left(\frac\ang3\right)
\sqrt{\left(\zfs + \qua - \az\right)^2 + 6\,\ax^2 + 3\,(\gyro + \gyroI)^2\,\field^2}\right],\\
\energy_5=&\, \frac32 \left[\zfs + \qua - \az + \cos\left(\frac{\ang + 2\uppi}3\right)
\sqrt{\left(\zfs + \qua - \az\right)^2 + 6\,\ax^2 + 3\,(\gyro + \gyroI)^2\,\field^2}\right],\\
\energy_6=&\, \frac23\left[\zfs + \qua - \az + \cos\left(\frac{\ang - 2\uppi}3\right)
\sqrt{\left(\zfs + \qua - \az\right)^2 + 6\,\ax^2 + 3\,(\gyro + \gyroI)^2\,\field^2}\right],\\
\energy_7 =&\, \frac12\left[\zfs + \qua - \left(\gyro - \gyroI\right)\field
+\sqrt{4\,\ax^2 + \left[\zfs - \qua - \left(\gyro + \gyroI\right)\field\right]^2}\right],\\
\energy_8 =&\, \frac12\left[\zfs + \qua - \left(\gyro - \gyroI\right)\field
-\sqrt{4\,\ax^2 + \left[\zfs - \qua - \left(\gyro + \gyroI\right)\field\right]^2}\right],\\
\energy_9 =&\, \az + \zfs + \qua - \left(\gyro - \gyroI\right)\field,
\end{align}
\end{subequations}

and the unnormal\IS ed eigenstates of \Equation~\eqref{eq:states-explicit}
expand to

\begin{subequations}\label{eq:14N-vec}
\begin{small}
\begin{align}
\ket{1} =&\, \ket{+1, +1},\\
    \nonumber\\
\ket{2} =&\, \ket{+1,0}\times\left(\left[\zfs - \qua + \left(\gyro + \gyroI\right)\field\right]
    + \sqrt{4\,\ax^2 + \left[\zfs - \qua + \left(\gyro + \gyroI\right)\field\right]^2} \right)\nonumber\\
    &+ \ket{0,+1}\times\sqrt{2},\\
    \nonumber\\
\ket{3} =&\, \ket{+1,0}\times\left(\left[\zfs - \qua + \left(\gyro + \gyroI\right)\field\right]
    - \sqrt{4\,\ax^2 + \left[\zfs - \qua + \left(\gyro + \gyroI\right)\field\right]^2} \right)\nonumber\\
    &+ \ket{0,+1}\times\sqrt{2},\\
    \nonumber\\
\ket{4} =&\,\ket{+1,-1}\times\bigg(
    4\,\left[\left(\zfs + \qua - \az\right)^2 + 6\,\ax^2 + 3\,(\gyro + \gyroI)^2\,\field^2\right]\cos^2\left(\frac{\ang}3\right)\nonumber\\
    &\quad\quad\quad\quad\quad\quad+2\,\left(\zfs + \qua - \az + 3\,\left(\gyro + \gyroI\right)\,\field\right)\,\sqrt{\left(\zfs + \qua - \az\right)^2 + 6\,\ax^2 + 3\,(\gyro + \gyroI)^2\,\field^2}\,\cos\left(\frac{\ang}3\right)\nonumber\\
    &\quad\quad\quad\quad\quad\quad+6\,\left[\zfs + \qua - \az\right]\left[\gyro + \gyroI\right]\,\field-9\,\ax^2 - 2\,\left[\zfs + \qua - \az\right]^2 \bigg)\nonumber\\
    &+\ket{0,0}\times3\,\ax\,\bigg[\az -\zfs - \qua + 3\,\left(\gyro + \gyroI\right)\,\field
    + 2\,\sqrt{\left(\zfs + \qua - \az\right)^2 + 6\,\ax^2 + 3\,(\gyro + \gyroI)^2\,\field^2}\,\cos\left(\frac{\ang}3\right)\bigg]\nonumber\\
    &+ \ket{-1, +1}\times9\,\ax^2,
\end{align}
\begin{align}
\ket{5} =&\,\ket{+1,-1}\times\bigg(
    4\,\left[\left(\zfs + \qua - \az\right)^2 + 6\,\ax^2 + 3\,(\gyro + \gyroI)^2\,\field^2\right]\cos^2\left(\frac{\ang + 2\uppi\,\rootIndex}3\right)\nonumber\\
    &\quad\quad\quad\quad\quad\quad+2\,\left(\zfs + \qua - \az + 3\,\left(\gyro + \gyroI\right)\,\field\right)\,\sqrt{\left(\zfs + \qua - \az\right)^2 + 6\,\ax^2 + 3\,(\gyro + \gyroI)^2\,\field^2}\,\cos\left(\frac{\ang + 2\uppi\,\rootIndex}3\right)\nonumber\\
    &\quad\quad\quad\quad\quad\quad+6\,\left[\zfs + \qua - \az\right]\left[\gyro + \gyroI\right]\,\field-9\,\ax^2 - 2\,\left[\zfs + \qua - \az\right]^2 \bigg)\nonumber\\
    &+\ket{0,0}\times3\,\ax\,\bigg[\az -\zfs - \qua + 3\,\left(\gyro + \gyroI\right)\,\field
    + 2\,\sqrt{\left(\zfs + \qua - \az\right)^2 + 6\,\ax^2 + 3\,(\gyro + \gyroI)^2\,\field^2}\,\cos\left(\frac{\ang + 2\uppi\,\rootIndex}3\right)\bigg]\nonumber\\
    &+ \ket{-1, +1}\times9\,\ax^2,\\
    \nonumber\\
\ket{6} =&\,\ket{+1,-1}\times\bigg(
    4\,\left[\left(\zfs + \qua - \az\right)^2 + 6\,\ax^2 + 3\,(\gyro + \gyroI)^2\,\field^2\right]\cos^2\left(\frac{\ang - 2\uppi\,\rootIndex}3\right)\nonumber\\
    &\quad\quad\quad\quad\quad\quad+2\,\left(\zfs + \qua - \az + 3\,\left(\gyro + \gyroI\right)\,\field\right)\,\sqrt{\left(\zfs + \qua - \az\right)^2 + 6\,\ax^2 + 3\,(\gyro + \gyroI)^2\,\field^2}\,\cos\left(\frac{\ang - 2\uppi\,\rootIndex}3\right)\nonumber\\
    &\quad\quad\quad\quad\quad\quad+6\,\left[\zfs + \qua - \az\right]\left[\gyro + \gyroI\right]\,\field-9\,\ax^2 - 2\,\left[\zfs + \qua - \az\right]^2 \bigg)\nonumber\\
    &+\ket{0,0}\times3\,\ax\,\bigg[\az -\zfs - \qua + 3\,\left(\gyro + \gyroI\right)\,\field
    + 2\,\sqrt{\left(\zfs + \qua - \az\right)^2 + 6\,\ax^2 + 3\,(\gyro + \gyroI)^2\,\field^2}\,\cos\left(\frac{\ang - 2\uppi\,\rootIndex}3\right)\bigg]\nonumber\\
    &+ \ket{-1, +1}\times9\,\ax^2,\\
    \nonumber\\
\ket{7} =&\, \ket{-1,0}\times\left(\left[\zfs - \qua - \left(\gyro + \gyroI\right)\field\right]
    + \sqrt{4\,\ax^2 + \left[\zfs - \qua - \left(\gyro + \gyroI\right)\field\right]^2} \right)\nonumber\\
    &+ \ket{0,-1}\times\sqrt{2},\\
    \nonumber\\
\ket{8} =&\, \ket{-1,0}\times\left(\left[\zfs - \qua - \left(\gyro + \gyroI\right)\field\right]
    - \sqrt{4\,\ax^2 + \left[\zfs - \qua - \left(\gyro + \gyroI\right)\field\right]^2} \right)\nonumber\\
    &+ \ket{0,-1}\sqrt{2},\\
    \nonumber\\
\ket{9} =&\, \ket{-1, -1},
\end{align}
\end{small}
\end{subequations}
where,
\begin{align}
\ang =&\, \arccos\left(\left[\az - \zfs - \qua\right]
    \frac{\left(\zfs + \qua - \az\right)^2 + 9\,\ax^2 - 9\,(\gyro + \gyroI)^2\,\field^2}{\left[\left(\zfs + \qua - \az\right)^2 + 6\,\ax^2 + 3\,(\gyro + \gyroI)^2\,\field^2\right]^{3/2}}\right).
\end{align}
\end{widetext}
\end{appendix}


\begin{thebibliography}{31}
\bibitem [{typ()}]{typical-nv}%
  \BibitemOpen
  \href@noop {} {}\bibinfo {note} {We are working on producing a set of
  world-average parameters for the \NV~\cite{typical-nv-real}. Here we list our
  current averages from the resources that we have currently found in the
  literature.}\BibitemShut {Stop}%
\end{thebibliography}

%

\end{document}